\documentclass[twocolumn,10pt]{asme2ej}

\usepackage{graphicx}
\usepackage{amsmath,amssymb}

\newcommand\Rey{\mbox{\textit{Re}}}  % Reynolds number

\title{Relaminarization of pipe flow by means of 3d-printed shaped honeycombs}

\author{Jakob K\"{u}hnen\thanks{Address for correspondence: jakob.kuehnen@ist.ac.at}, Davide Scarselli, Bj\"{o}rn Hof \\
    \affiliation{
	IST Austria\\
	Am Campus 1\\
	A-3400, Klosterneuburg\\
    }	
}

\graphicspath{{./figure/}}

\begin{document}

\maketitle    

%%%%%%%%%%%%%%%%%%%%%%%%%%%%%%%%%%%%%%%%%%%%%%%%%%%%%%%%%%%%%%%%%%%%%%
\begin{abstract}
	%An abstract for an ASME paper should be less than 150 words and is normally in italics.
{\it Based on a novel control scheme, where a steady modification of the streamwise velocity profile leads to complete relaminarization of initially fully turbulent pipe flow, we investigate the applicability and usefulness of custom-shaped honeycombs for such control. The custom-shaped honeycombs are used as stationary flow management devices which generate specific modifications of the streamwise velocity profile. Stereoscopic particle image velocimetry and pressure drop measurements are used to investigate and capture the development of the relaminarizing flow downstream these devices. We compare the performance of straight (constant length across the radius of the pipe) honeycombs with custom-shaped ones (variable length across the radius). An attempt is made to find the optimal shape for maximal relaminarization at minimal pressure loss. The maximum attainable Reynolds number for total relaminarization is found to be of the order of 10.000. Consequently the respective reduction in skin friction downstream of the device is almost by a factor of 5. The break-even point, where the additional pressure drop caused by the device is balanced by the savings due to relaminarization and a net gain is obtained, corresponds to a downstream stretch of distances as low as approx.\ 100 pipe diameters of laminar flow.
}
\end{abstract}

%%%%%%%%%%%%%%%%%%%%%%%%%%%%%%%%%%%%%%%%%%%%%%%%%%%%%%%%%%%%%%%%%%%%%%
%\begin{nomenclature}
%\entry{A}{You may include nomenclature here.}
%\entry{$\alpha$}{There are two arguments for each entry of the nomemclature environment, the symbol and the definition.}
%\end{nomenclature}
%%%%%%%%%%%%%%%%%%%%%%%%%%%%%%%%%%%%%%%%%%%%%%%%%%%%%%%%%%%%%%%%%%%%%%

\section{Introduction}

%mostly copied from JFM-FMD-Control_Rev3.tex

As recently shown by K\"{u}hnen et al.\ \cite{Kuehnen2018a} a suitable steady modification of the streamwise velocity profile in a pipe can lead to a complete collapse of turbulence and the flow fully relaminarizes. The necessary modification of the profile was shown to be such that the resulting streamwise velocity profile was (more) plug shaped and flat or even dented in the center of the pipe, requiring that the flow in the near wall region is accelerated and the flow in the center of the pipe is decelerated as compared to the uncontrolled velocity profile in a pipe. In their numerical simulations \cite{Kuehnen2018a} simply added an appropriate radially dependent body force term to the equation of motion, modifying the streamwise velocity profile to a more plug-like one. In experiments, several practical techniques to modify the mean velocity profile of turbulent flow were shown to be feasible. One technique was by stirring the flow with four rotors. Another approach, elaborated in \cite{Scarselli2018}, was by means of a movable pipe segment which was used to locally accelerate the flow at the wall. A further approach, elaborated in \cite{Kuehnen2018b}, was by injecting fluid through an annular gap at the wall to accelerate the flow close to the wall or by inserting an obstacle partially blocking the pipe. All approaches force the streamwise velocity profile in a similar manner and can modify the streamwise velocity profile appropriately to achieve relaminarization. The experiments demonstrated that relaminarization occurred as a direct result of a particular streamwise velocity profile which exhibited a severely decreased lift-up potential that can not sustain turbulence.

In the present investigation we want to explore a further possibility of specifically modifying the streamwise velocity profile employing passive flow management devices. The aim is to modify the streamwise velocity profile in a pipe similar to \cite{Kuehnen2018b} by solely passive means to achieve complete relaminarization of initially fully turbulent pipe flow. 
%When thinking of possible passive devices which could represent a deliberate obstacle or obstruction acting as a cross-sectionally extended volume forcing one necessarily ends up with considerations of screens and honeycombs. 
To homogeneously force the streamwise velocity profile in a pipe a very fine meshed screen with radially varying mesh size may seem the simplest choice, however unwanted apart from modifications of the streamwise velocity such screens also caused cross stream components which can prevent relaminarization. It is clear that any physical body inserted into the flow, be it as small as technically feasible, can reduce the level of existing turbulence and control the flow profile -- but it can at the same time also produce flow inhomogeneities and turbulent structures.
%Flattening the streamwise profile in a pipe can in principle be achieved by either decelerating the flow at the pipe center or by accelerating the fluid at the wall.

Screens have been used for decades to reduce or suppress turbulence, to break down incoming large scale structures and to make the flow more uniform in all kinds of applications such as wind- and watertunnels. The various types of screens are commonly referred to as grid, mesh, sieve, woven wire, gauze or, if the device is markedly extended in the streamwise direction, honeycomb. A further distinction can then be made between real honeycombs (regular hexagonal cells) and other types of cells, e.g.\ square cells or circular cells like in devices made of drinking straw \cite{Mikhailova1994}. Screens can be generally seen as flow conditioners which damp or even extinguish the incoming, upstream turbulence and largely diminish velocity or pressure nonuniformities. The suppression of turbulence is mostly due to the inhibition of lateral (transverse) components of the flow, obtained at the cost of an additional pressure drop across the screen. In any application this penalty is to be weighed against the positive effect of turbulence reduction. Since screens can also generate, primarily through documented instabilities, new turbulence with scales characteristic of the shear layers present in their near wake downstream, the control of turbulence by means of screens and honeycombs is always a balance between suppression and generation \cite{Loehrke1976}.

Concise discussions concerning suppression of turbulence by screens can be found in e.g.\ \cite{Bradshaw1964}, \cite{Laws1978}, \cite{Groth1988} and \cite{Oshinowo2000} and references therein. The honeycomb type of screens, which are generally useful for turbulence reduction if swirl or initially high transverse velocities are present, is discussed in e.g.\ \cite{Loehrke1976}, \cite{Lumley1967}, \cite{Farell1996} and \cite{Kulkarni2011}. However, the data available on honeycombs is rather limited and mainly given as best practice and empirical laws from experiments. The usage of screens to increase turbulence levels or to generate quasi-isotropic turbulence \cite{Roach1987,Liu2007}, often used to investigate dissipation and naturally decaying free-stream turbulence downstream the screen \cite{Oshinowo2000,Valente2011,Vassilicos2015}, is an operational mode which will not be considered further in the present context.

\cite{Kotansky1966,Livesey1973,Sajben1975,Ahmed1997} have successfully used screens and honeycombs to produce artificial, more or less well controlled velocity profiles in free-stream turbulence. They demonstrated that the velocity profile downstream of screens or honeycombs can be controlled via a variation of the mesh size or the cell length. Since in the present investigation the variation of the mesh size is limited by the capabilities of the 3d-printer (see section \ref{subsec:fmd}), the present work is confined to varying the cell length of regular hexagonal honeycombs across the cross-section of the pipe. I.e., we employ shaped honeycombs to generate a particular profile of the streamwise velocity. The success of the relaminarization devices in our study predominantly appears to depend on the shape of the generated streamwise velocity profile. The suppression of turbulence due to the inhibition of transverse components certainly also plays a role but to a lesser extend. As an attempt to distinguish the effect of the forcing on the streamwise profile from normal turbulence suppression through the annihilation of transverse fluctuations we also present results of unshaped honeycombs, where all cell lengths are equal. 

The outline of the paper is as follows. In the next section we describe the experimental facility and the custom 3d-printed honeycomb-devices which were used to generate velocity profiles. Then we describe the examination procedure and the steps to systematically find the optimal flow profile and produce a specified profile with the least effort in terms of pressure drop. The results are presented and discussed in section \ref{sec:results}.

\section{Experimental setup and method}\label{sec:experimenalsetup}

The test facility consists mainly of a straight long glass-pipe with turbulent flow. 3d-printed honeycomb devices act as flow management devices (FMDs) which can be mounted within the pipe between two pipe sections. The FMDs form a deliberate obstacle or obstruction acting as a spatially extended volume forcing on the flow. The facility allows to investigate the effect of the FMDs on the flow. In the following section the facility and the FMDs are described in detail.
 
\subsection{Facility}\label{subsec:facility}

\begin{figure} %width=0.94\textwidth to be changed
	\centerline{\includegraphics[clip,width=0.48\textwidth,angle=0]{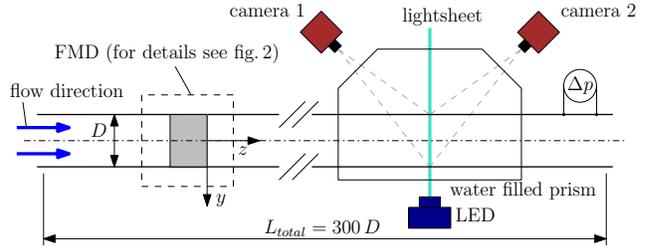}}
	\caption{\label{fig:sketch-setup_obst}Sketch of the test facility. Different flow management devices (see fig.\ \ref{fig:sketch_hc-fmd} for details) can be mounted inside a glass pipe. The flow direction is from left to right. Drawing not to scale.}
\end{figure}

Figure \ref{fig:sketch-setup_obst} shows a sketch of the test facility used to test different FMDs. The setup consists of a glass-pipe with inner diameter $D=30\pm0.01$\,mm and a total length of 9\,m ($300\,D$) made of 1 meter sections. Water driven by gravity enters the pipe from a reservoir located 20\,m above the pipe. The flow rate and hence the Reynolds number ($Rey= U D/{\nu}$ where $U$ is the mean velocity, $D$ the diameter of the pipe and $\nu$ the kinematic viscosity  of the fluid) can be adjusted by means of a control valve in the supply pipe.
%To ensure that always fully turbulent flow enters the measurement section even at moderate Reynolds numbers, a tiny steady perturbation (needle) is placed $10\,d$ after the entrance.

The velocity field is measured $\sim250\,D$ downstream from the inlet at the position of the lightsheet. The measurement plane is perpendicular to the streamwise flow direction (pipe $z-$axis). All three velocity components within the plane are recorded using a high-speed stereo PIV system (Lavision GmbH) consisting of a pulsed LED (IL-106X LED Illuminator, HARDsoft Microprocessor Systems) and two Phantom V10 high-speed cameras with a full resolution of $2400\times 1900$\,px. Vestosint particles (mean diameter 45\,$\mu$m, $\rho=1.016$\,g/cm$^3$, Evonik Degussa GmbH) were used as seeding particles. We use a local non-dimensional Cartesian coordinate-system ($x,y,z$) = ($\tilde{x}/D$, $\tilde{y}/D$, $\tilde{z}/D$) as indicated in figure \ref{fig:sketch-setup_obst}, where the downstream end of the FMD is always located at the origin. The respective Cartesian velocity components ($\tilde{u},\tilde{v},\tilde{w}$) are made non-dimensional by dividing them by the mean velocity $U$, yielding ($u,v,w$). Around the measurement plane the pipe is encased by a water filled prism such that the optical axes of the cameras are perpendicular to the air--water interface to reduce refraction and distortion of the images.

A differential pressure sensor (DP\,103, Validyne) is used to measure the pressure drop $\Delta p$ between two pressure taps. As turbulent flows experience a skin drag much higher than the laminar one and in addition, the difference becomes larger as the Reynolds number increases, we use $\Delta p$ for a straightforward detection of the flow state based on the large difference between turbulent and laminar friction factors ($f_{\mathrm{turb}}$ and $f_{\mathrm{lam}}$). For this purpose the sensor is placed at the downstream end of the pipe as indicated in fig.\ \ref{fig:sketch-setup_obst}, and the pressure taps are separated by $30\,D$ in the streamwise direction. 
%The friction factor is defined as
%\begin{equation}
%f = \frac{8\tau_w}{\rho U_b^2},
%\end{equation}
%where $\tau_w$ is the mean shear stress at the pipe wall. 
For $Re\lesssim80\,000$ the ratio of $f_{\mathrm{turb}}$ and $f_{\mathrm{lam}}$ can be expressed with the aid of Blasius correlation \cite{Pope2001}
\begin{equation}\label{eq.ratiolamturb}
\frac{f_{\mathrm{lam}}}{f_{\mathrm{turb}}} = \frac{64/Re}{0.316/Re^{0.25}}\sim Re^{-0.75},
\end{equation}
which also shows that even at low Reynolds numbers relaminarizing a turbulent flow can produce a large drag reduction. %For example, already at $\Rey=3000$ we already have a gain of about 50\%.

Furthermore, we measure $\Delta p_L$ across the FMDs to compute their pressure drop coefficients $K$ according to
\begin{equation}\label{eq.K}
K = \frac{2\Delta p_L}{\rho U^2 }.
\end{equation}
where $\rho$ is the fluid density. For this purpose the pressure taps are separated by $12\,D$ in the streamwise direction and the FMD is placed $2\,D$ downstream the upper tap. 
%This relatively short measuring distance is chosen because the flow upstream the FMD is turbulent, while the flow downstream the FMD is supposed to be mostly laminar. A longer distance, commonly recommend to ensure a sufficient development length, did not seem appropriate in our case.

% $\Delta p=K\,0.5 \rho U^2 $,
%where $K$ is the (static) pressure drop coefficient. For the dependence of the static pressure drop coefficient on the Reynolds number and solidity / porosity: see again \cite{Groth1988}, page 142.

\subsection{Flow management device}\label{subsec:fmd}

In order to control the flow in a way that the streamwise velocity profile becomes more flat a stationary FMD must redirect the flow such that the velocity is decelerated in the center of the pipe and accelerated close to the wall as compared to the uncontrolled velocity profile. We employ shaped honeycombs with variable length across the cross-section of the pipe to generate a particular profile of the streamwise velocity.

\begin{figure}
	\centerline{\includegraphics[clip,width=0.48\textwidth,angle=0]{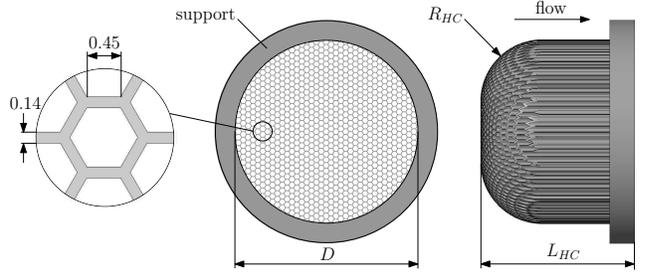}}
	\caption{\label{fig:sketch_hc-fmd}Front view and side view of the FMD. The support is mounted within a flange to fix the FMD within the pipe. All dimensions in mm.}
\end{figure}

Rapid prototyping by means of a 3d-printer (ProJet 3510 HD, 3D Systems, Inc.) is used to produce the FMDs as shown in fig.\ \ref{fig:sketch_hc-fmd}. These FMDs consist of regular hexagonal honeycombs. The support (retaining ring) is used to keep the FMD between two pipe sections within a flange connector of the pipe. The cells of the honeycomb have a side length of $l=0.45$\,mm and a wall thickness of $0.14$\,mm as shown in the detail of the figure on the left. The equivalent (hydraulic) diameter of a single cell is $d_h=4A/P=0.78$\,mm, where $A$ is the cross sectional area and $P$ the perimeter of the cell. The porosity of the honeycomb, i.e.\ the ratio of open to blocked area, is $\beta=71.8\%$.

The FMDs are printed with different total lengths ($L_{HC}$) and different overall shapes, i.e.\ with radially varying cell length. The cell length is varied in the radial direction by beveling the FMD with a varying radius $R_{HC}$ from $0$ to $14$\,mm in the maximum case. The flow direction through the shaped FMDs is as indicated in fig.\ \ref{fig:sketch_hc-fmd}, i.e.\ the straight side facing downstream. Furthermore, several unshaped honeycombs (where $R_{HC}=0$) with $L_{HC}=1,3,5,7,10,15$ and $20$\,mm, yielding relative lengths $L_{HC}/d_h=1.3-25.6$, were investigated for comparison. For reference in the text each particular FMD is assigned a shortcut following the naming convention FMD-$L_{HC}$-$R_{HC}$.

\begin{figure}
	\centerline{\includegraphics[clip,width=0.46\textwidth,angle=0]{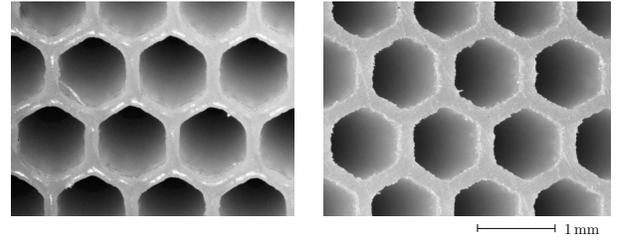}}
	\caption{\label{fig:hc-front-backside}Magnified images of the frontside (left) and backside (right) of the printed honeycomb. The backside had to be sanded after printing. The length of the arrow indicates the scale.}
\end{figure}

The whole device including the support is printed in one piece as a precise, durable plastic part. The production drawing of the FMD and a .stl-file which can be used to print the FMDs with a 3D-printer is provided in the online supporting material. All tested FMDs are printed in UHD mode (resolution $750\times750\times890$\,DPI, specified accuracy of 0.025-0.05\,mm per 25.4\,mm) using the material VisiJet Crystal. Figure \ref{fig:hc-front-backside} (left) shows a magnified image of the frontside of a printed honeycomb. The overall dimensional precision can be considered very high, although small deviances are visible especially in the corners of the cells. Pretests have shown that wall thicknesses $<0.14$\,mm, although desirable in the present context, could not be reproduced reliably. Due to the printing process the first layer of the print, i.e.\ where the printer starts to print the device, is slightly rougher and contaminated with wax which is used as support material in the printing process. This side, which was usually the flat side of the print and hence the backside of the FMD, needed to be sanded after printing. Figure \ref{fig:hc-front-backside} (right) shows that even sanding with very fine sand paper left tiny burrs at the front edges of the honeycomb. 

To investigate the consequences of a modified streamwise velocity profile we characterize the effect of the FMDs by measuring the modification of the mean velocity distribution and the flow development downstream of the screen as well as the pressure drop across the FMDs. Determining the success of the modification is quite simple. If we observe a laminar pressure drop and a parabolic profile further downstream of the device the modification of the velocity profile is regarded effective. After identifying the optimal profile for maximum relaminarization, i.e.\ at the highest Reynolds number possible, we try to optimize the device so that the pressure drop is minimized.

In a number of references the importance of having tight tolerances has been pointed out \cite{Roach1987}. Deviations in dimensional accuracy cause variations of the pressure-drop coefficient from point to point and can produce large uncontrolled variations in the downstream velocity profile. From our measurements we can confirm that slight imperfections can have a profound influence on the downstream flow characteristics. Already a single cell being partially blocked by an air bubble or dirt can prevent relaminarization.

\section{Results and discussion}\label{sec:results}

In this section we present the results of stereo PIV and pressure drop measurements and discuss the outcome. However, to get a quick overview all FMDs were first investigated by means of mere visualization. Neutrally buoyant anisotropic particles \cite{Matisse1984} were added as tracer particles and the flow in the pipe was illuminated by means of LED string lights along the whole length of the pipe to be able to observe the development of the flow field up- and downstream the FMDs (similar to \cite{Kuehnen2015video} and \cite{Kuehnen2017video}). %Using this procedure we could reliably detect air bubbles etc.\ and if the flow is laminar or turbulent.

\begin{figure}
	\centering
	\includegraphics[clip,width=0.46\textwidth,angle=0]{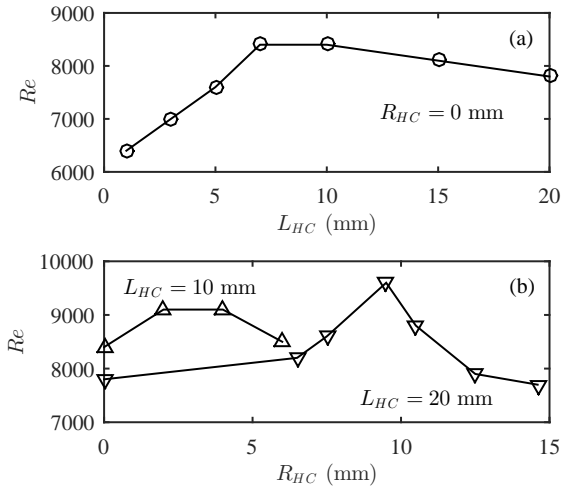}
	\caption{Maximum values of $\Rey$ at which full relaminarization is observed for (a)  straight honeycombs ($R_{HC}=0$\,mm) and (b) two selected examples ($L_{HC}=10$\,mm and $L_{HC}=20$\,mm) of radially shaped ones as a function of $R_{HC}$.}
	\label{fig:max_Re}
\end{figure}

We varied the Reynolds number in increments of 100 and observed either laminar or turbulent flow downstream the FMDs. Figure \ref{fig:max_Re} (a) displays the maximum values for which laminar flow was found (i.e.\ we measured $f_{\mathrm{lam}}$) for straight ($R_{HC}=0$\,mm) honeycombs with lengths from $L_{HC}=1$ to 20\,mm. Already the shortest honeycomb with $L_{HC}=1$\,mm exhibits an unexpectedly high capability of relaminarizing the flow. Up to $\Rey=6400$ the flow downstream this honeycomb is found to be laminar and stay laminar for the remainder of the pipe. By increasing the length of the FMD up to 10\,mm the flow can be made laminar even up to $\Rey=8400$. A further increase in length does not seem beneficial anymore, as the maximum values for relaminarization decrease. In a next step we added a radial shape to the FMDs and tested the devices again in increments of 100. Figure \ref{fig:max_Re} (b) depicts the relaminarizing capability for two selected examples ($L_{HC}=10$\,mm and $L_{HC}=20$\,mm) of radially shaped FMDs (as a function of $R_{HC}$). The radial shaping can improve the relaminarizing capability of the FMDs even further. The FMD with $L_{HC}=10$\,mm reaches $\Rey=9100$ with $R_{HC}=2-4$, the FMD with $L_{HC}=20$\,mm even reaches $\Rey=9600$ with $R_{HC}=9.5$. Again, a further increase in $R_{HC}$ does not seem beneficial anymore, as the maximum values for relaminarization decrease with further increasing $R_{HC}$.

\begin{figure}
	\centerline{\includegraphics[clip,width=0.46\textwidth,angle=0]{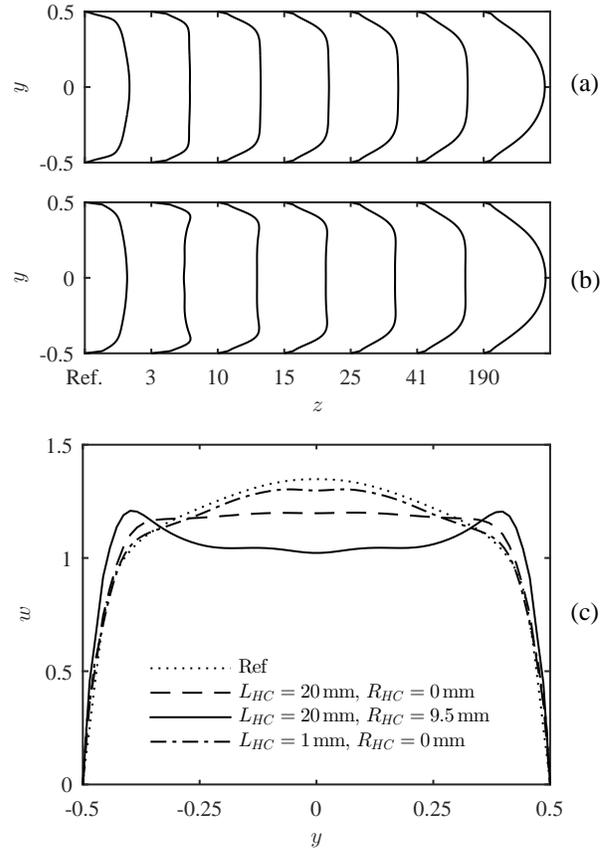}}
	\caption{Downstream evolution of mean streamwise velocity profiles at $\Rey=6000$ for (a) $L_{HC}=20$\,mm and $R_{HC}=0$\,mm and (b) $L_{HC}=20$\,mm and $R_{HC}=9.5$\,mm. In both plots the profile of uncontrolled turbulent flow is shown for reference (Ref) at the left. (c) provides a quantitative comparison of selected profiles at $z=3$.}
	\label{fig:Evolution}
\end{figure}

To characterize the flow downstream the FMDs we took PIV-measurements at several locations along $z$, with $z=3$ being the closest possible downstream location. Figure \ref{fig:Evolution} (a) and (b) exemplary show the appearance and evolution of the mean streamwise velocity profile downstream a straight FMD ($L_{HC}=20$\,mm) and one with an additional radial shape ($R_{HC}=9.5$\,mm). In both of the two representative cases the FMDs similarly relaminarize the flow at the given Reynolds number, verified by the evolution of the profiles towards a parabolic shape (clearly manifest for the profiles at $z=190$). However, there is a particularly noteworthy difference in the time-averaged velocity profiles right downstream the FMDs, visible for the profiles at $z=3$. While the profile for the straight FMD is clearly flattened as compared to the reference flow, the profiles for the shaped FMD exhibit an M-shaped appearance.

%XXX Difference between equally flat profiles is decreasing turbulent kinetic energy, see plot tke re=6000 (not shown). XXX

To enable a quantitative comparison of the flattening due to the straight FMD and the M-shape due to the shaped FMD fig.\ \ref{fig:Evolution} (c) depicts four different velocity profiles at $z=3$. The reference profile is measured in uncontrolled turbulent flow. The velocity profile for FMD-1-0 ($L_{HC}=1$\,mm, $R_{HC}=0$\,mm) exhibits a slight yet not very pronounced flattening in comparison. For the much longer FMD-20-0 the velocity profile is almost perfectly flat in the central area of the cross-section. The radially shaped FMD-20-9.5 produces an M-shaped appearance. The M-shape is caused by overshoots of faster fluid relatively close to the wall (peak at $y\pm0.4$) and a clearly decreased centerline velocity ($\sim1$) causing the M-shaped appearance with a pronounced plateau in the center. 

Interestingly, this M-shaped streamwise velocity profile is the one that provides the best relaminarizing capability we observed. According to fig.\ \ref{fig:max_Re} (b) FMD-20-9.5 can relaminarize a flow up to $\Rey=9600$, which is 23$\%$ above the unshaped FMD-20-0 and  14$\%$ above the best working unshaped FMD-10-0. Although those absolute values should be treated with caution and might vary for different FMDs (with differently sized honeycombs etc.) and different pipe diameters, the trend that the M-shape is advantageous compared to a mere flattening of the profile is consistent with observations of \cite{Scarselli2018} and \cite{Kuehnen2018b}. 

According to the findings reported in \cite{Kuehnen2018a} a flattened streamwise velocity profile relaminarizes because it exhibits a severely decreased lift-up potential \cite{Brandt2014} and thus can not sustain the turbulence regeneration cycle. As a measure for the reduced amplification mechanism of the regeneration cycle they consider the linearized Navier-Stokes equations and perform a transient growth (TG) analysis (following the algorithm given by \cite{Butler1993}, for further details see also \cite{Kuehnen2018a} and \cite{Meseguer2003}). The velocity profiles of all successfully controlled, i.e.\ relaminarizing flows considered by \cite{Kuehnen2018a}, \cite{Scarselli2018} and \cite{Kuehnen2018b} are shown to have a substantially reduced transient growth. We applied the same procedure here to the exemplary velocity profiles shown in fig.\ \ref{fig:Evolution} (c). We find a value of 160 for the uncontrolled reference flow, 121 for FMD-1-0, 68.8 for FMD-20-0 and 39.4 for FMD-20-9.5. In other words, the profiles consistently show a considerably decreasing TG with increasing length of the FMDs and even more for the shaped FMDs. The mere flatting of the velocity profile due to FMD-20-0 is inferior to FMD-20-9.5, which produces an M-shaped streamwise velocity profile with the best relaminarizing capability and the lowest transient growth. %However, $z=3$ is not necessarily the streamwise location of the minimum transient growth. More refined measurements would be necessary to find the axial location of the minimum TG along $z$.

%\begin{center}
%	XXX Values of transient growth of the profiles\\
%	Reference: 116.8\\
%	$R_{HC}=0$\,mm: 68.8\\
%	$R_{HC}=9.5$\,mm: 39.3\\
%	XXX
%\end{center}

%%%%%%%%%%%%%%%%%%%%%%%%%%%%%%
\begin{figure}
	\centerline{\includegraphics[clip,width=0.46\textwidth,angle=0]{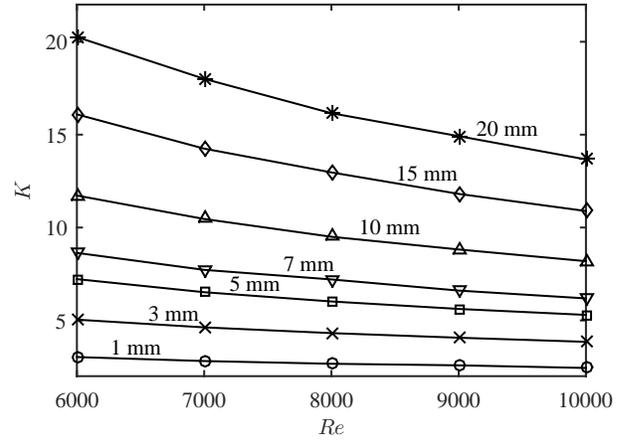}}
	\caption{\label{fig:re-k} Pressure drop coefficient $K$ as a function of $\Rey$ for different straight FMDs ($R_{HC}=0$\,mm). Each curve is labelled by $L_{HC}$.}
\end{figure}

In a next step we measure the pressure drop $\Delta p$ across the FMDs with length $1,3,5,7,10,15$ and 20\,mm and compute the pressure drop coefficient $K$ according to eq.\ \ref{eq.K}. Figure \ref{fig:re-k} shows $K$ in the range $6000\leq\Rey\leq10000$. As to be expected, longer FMDs result in a larger normalized pressure drop and overall this decreases with the Reynolds number. Our results show a trend in good agreement with the data reported in \cite{Loehrke1976}, although here the authors used honeycombs made of plastic straws at considerably higher Reynolds numbers.

\begin{figure}
	\centering
	\includegraphics[clip,width=0.46\textwidth,angle=0]{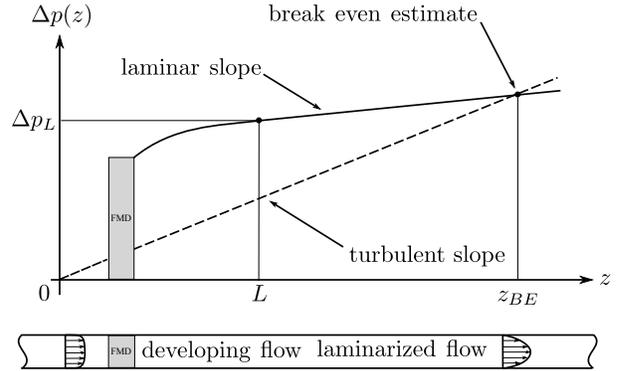}
	\caption{Sketch of the qualitative evolution of the pressure drop $\Delta p(z)$ along the pipe. In the undisturbed flow $\Delta p(z)$ increases linearly with $z$ with a slope given by the turbulent friction factor (dashed line). The presence of relaminarizing honeycomb-FMD (grey rectangle) introduces an abrupt increase of $\Delta p(z)$. Further downstream the flow develops to laminar and $\Delta p(z)$ grows linearly with a slope proportional to the laminar friction factor (solid line). The intersection between the two curves is the break-even point and represents the minimum length of pipe necessary to realize an energy gain.}
	\label{fig.estimate-norm}
\end{figure}

%by davide
Modifying the flow by means of a passive obstacle such as the FMDs comes with the additional cost of a concentrated pressure loss. Thus, the actual realization of a net energy saving for transporting the fluid can be achieved only when the gain due to the laminar pressure drop exceeds the concentrated loss. Figure \ref{fig.estimate-norm} shows the qualitative behavior of the pressure drop $\Delta p (z)$ with respect to the pressure tap located 2$D$ upstream the honeycomb-FMD. The solid and dashed lines qualitatively represent the flow with and without the FMD, respectively. The presence of the obstacle results in an abrupt jump of the pressure. However, if relaminarization actually takes place, then at some point downstream the two lines intersect each other.

The distance of such an energetic break-even point from the FMD is a suitable measure for characterizing the performance of the FMDs. Generally, it depends on both the Reynolds number and on the concentrated pressure loss introduced by the FMD. To estimate the distance of the break-even point $z_{BE}$ we look for the intersection between $\Delta p(z)$ without FMD (dashed line, fig.\ \ref{fig.estimate-norm}) and $\Delta p(z)$ with the FMD (solid line). The distance is thus given by
\begin{equation}
z_{BE} = \frac{\Delta p_L - Lf_{\mathrm{lam}}}{f_{\mathrm{turb}}-f_{\mathrm{lam}}}
\label{eq:BE}
\end{equation}

%An accurate localization of the break-even point requires the knowledge of $\delta p(z)$  to find the intersection with the turbulent curve. From an experimental standpoint, this would require many pressure measurements at various $z$ stations and a pipe sufficiently long to get the fully developed laminar flow. Moreover, the break even point is presumably highly dependent on the specific FMD and the necessary pipe length is not known in advance. To overcome the aforementioned issues we follow a different approach. Instead of trying to directly measure the distance of the break-even point we look for a lower-bound estimate. Assuming a laminar trend for $2\delta p(z)$ we compute the lower bound estimate of the break-even point as

\begin{figure}
	\centering
	\includegraphics[clip,width=0.46\textwidth,angle=0]{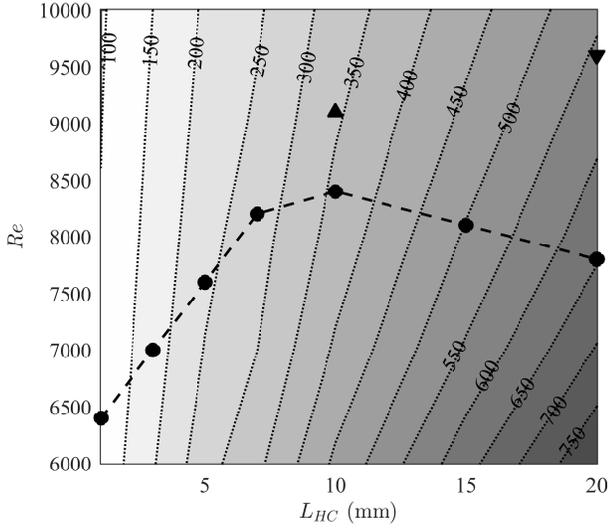}
	\caption{Contour levels of the break-even location $z_{BE}$ (in pipe diameter) estimated based on eq.\ \eqref{eq:BE} for straight FMDs ($R_{HC}=0$ mm). The bullet-points $\bullet$ represent the maximum value of $\Rey$ at which relaminarization is observed (see figure \ref{fig:max_Re}a). For $L_{HC}=10$\,mm ($\blacktriangle$) and $L_{HC}=20$\,mm ($\blacktriangledown$) we marked  the maximum value of $\Rey$ at which relaminarization is observed with shaped honeycombs (see figure \ref{fig:max_Re}b). }
	\label{fig:breakeven}
\end{figure}

%By using eq.\ \eqref{eq:BE} we estimate the distance from the FMD to the break-even point $z_{BE}$. 
The results are presented in fig.\ \ref{fig:breakeven} together with the curve which describes the maximum Reynolds number for which an FMD outputs a stable laminar flow. Notice that the estimate of $z_{BE}$ for $R_{HC}=0$\,mm is also a conservative estimate in case $R_{HC}\neq0$\,mm. As can be seen, FMD-1-0 can reach the energetic break-even point within almost $100\,D$ downstream. FMD-10-0 provides a considerably increased relaminarization capability, however, also $z_{BE}$ is shifted downstream to $\approx360\,D$. FMD-10-2, which relaminarizes the flow up to $\Rey=9100$, needs slightly less than $350\,D$ to reach $z_{BE}$. In other words, only if the pipe is longer than $350\,D$ a net energy gain can be achieved. FMD-10-9.5, which relaminarizes the flow up to the highest Reynolds number of 9600, needs $\sim550$ pipe diameter to reach $z_{BE}$. 

%According to eq.\ \ref{eq.ratiolamturb} this implies a reduction in skin friction by a factor of 5.

%\cite{Frohnapfel2012}: Under the constant flow rate (CFR) condition, a successful drag-reducing technique effectively reduces friction drag, which immediately translates into a reduction of the pumping energy. Since by definition active control requires additional energy, recent investigations are shifting their emphasis from the raw drag reduction rate R to the net energy saving rate S, which corresponds to the reduction rate of the total energy consumption, and also the gain G, defined as the ratio between the reduction of pumping energy and the control energy input.

%%%%%%%%%%%%%%%%%%%%%%%%%%%%%%%%%%%%%%%%%%%%%%%%%%%%%%%%%%%%%%%%%%%%%%
\section{Conclusions}

Employing 3d-printed honeycombs to control the flow we can not clearly distinguish between several parameters acting on the flow, as the suppression of transverse components and fluctuations in the flow through the honeycombs can annihilate turbulence without any further measures up to Reynolds numbers considerably higher than the critical Reynolds number of $\sim2040$ \cite{Avila2011} for sustained turbulence. However, our investigation has shown that custom-shaped honeycombs with variable length across the cross-section can be used to produce optimized streamwise velocity profiles for maximal relaminarization in a pipe. The maximally achievable Reynolds number for complete relaminarization we found is of the order of 10\,000. Although this is still a relatively low Reynolds number in terms of industrial applications, the achievable drag reduction is already by a factor of five according to eq.\ \ref{eq.ratiolamturb}. Our lower bound estimate for the distance, where a net energy gain can be achieved, is $100\,D$ for the shortest FMD-1-0 and $\sim550$ for the FMD-20-9.5 exhibiting the highest relaminarizing capability. The ideal streamwise velocity profile for relaminarization is shown to be not just a flattened profile, but to exhibit a specific M-shape, i.e.\ with velocity overshoots close to the wall and a plateau in the center of the pipe.

Several further geometries, in particular a radial variation of the cell-size, could be very interesting targets for future investigations and improvements in terms of pressure drop and maximal relaminarization. After all, pipe flow exhibits a feature that makes it particularly attractive for relaminarization methods: the laminar state is stable to infinitesimal perturbations at all flow speeds \cite{Drazin1981}. Consequently, once relaminarization is achieved, the flow remains laminar as long as the pipe is straight and smooth. Turbulence can only return if a sufficiently strong disturbance is encountered. 

%see again: \cite{Groth1988}
%see again sreenivasan1982,page 34, 7.3 Mean Parameters

%The pressure loss across uniform screens has been shown to be a function of screen solidity with no Reynolds number dependence for lo2 < Red < lo4, where the Reynolds number is calculated based on the diameter of the wire/rod/bar (Laws and Livesey, 1978). from oshinowo 2000, very similar also in roach1986

% see: http://navier.stanford.edu/bradshaw/tunnel/honeycomb.html

%%%%%%%%%%%%%%%%%%%%%%%%%%%%%%%%%%%%%%%%%%%%%%%%%%%%%%%%%%%%%%%%%%%%%%
%\section{Discussions}

%%%%%%%%%%%%%%%%%%%%%%%%%%%%%%%%%%%%%%%%%%%%%%%%%%%%%%%%%%%%%%%%%%%%%%
\begin{acknowledgment}
The project was partially funded by the European Research Council under the European Union’s Seventh Framework Programme (FP/2007-2013)/ERC grant agreement 306589. The authors declare that they have no conflict of interest. We thank M.\ Schaner and T.\ Asenov for valuable assistance in producing the FMDs.

\end{acknowledgment}

%%%%%%%%%%%%%%%%%%%%%%%%%%%%%%%%%%%%%%%%%%%%%%%%%%%%%%%%%%%%%%%%%%%%%%
% Here's where you specify the bibliography style file.
% The full file name for the bibliography style file 
% used for an ASME paper is asmems4.bst.
\bibliographystyle{asmems4}

% Here's where you specify the bibliography database file.
\bibliography{Literatur_JK}

\end{document}